\documentstyle[preprint,prb,aps]{revtex}

\tightenlines

\begin{document}
\title{{\em Ab initio} Hartree-Fock with electronic correlation study of the
electronic properties of MgB$_{2}$}
\author{Armando Reyes-Serrato\cite{ARS} and Donald H. Galv\'{a}n}
\address{Centro de Ciencias de la Materia Condensada de la UNAM\\
Apartado Postal 2681, Ensenada, Baja California, 22800 M\'{e}xico.}
\date{\today}
\maketitle

\begin{abstract}
We performed {\em all-electron ab initio} self-consistent field Hartree-Fock
linear combination of atomic orbital, in which electronic correlation using
density functional were included to perform electronic calculations in MgB$%
_{2}$ new superconductor. Superconductivity in this compound was correlated
with existence of p$_{x,y}$ band holes at $\Gamma $ point.
\end{abstract}


\section{INTRODUCTION}

MgB$_{2}$ has attracted lots of attention of the international scientific
community due that it shows a superconducting transition of the order of 39
K.\cite{1} Furthermore, its unitary cell (three atoms per unit cell), is
simple enough that provides a tool to understand the mechanism proposed\cite
{2} for this material to behave like a BCS\cite{3} type of superconductor 
{\em i.e.} electron-phonon interaction.

Early theoretical studies\cite{4,5} of the electronic structure, using other
methods of calculations like OPW and tight-binding, have shown similar
behavior to graphite, from the point of view of band theory. Transition
metal diborides (TMD), as expected, show a more complex structure due to the
consideration of d-states.

Recently, an electronic structure calculation on MgB$_{2}$\cite{6} shows
that some states of boron are responsible for the metallic behavior of this
compound. Furthermore, phonon frequencies and electron-phonon coupling (at
the most symmetric point of the Brillouin zone) lead these authors to
conclude to phonon based superconductivity.

\section{ COMPUTATIONAL DETAILS}

The calculations reported in the present work had been carried out using
CRYSTAL98\cite{7} {\em ab initio} Self-consistent-field (SCF) Hartree-Fock
linear combination of atomic orbital computer program which provide solution
to crystalline system of any space group. Electronic correlation using
density functional (DFT) has been included as a correction to the total
energy using the correlation function proposed by Perdew-Zunger (PZ).\cite{8}
More details about the mathematical formulation of CRYSTAL98 have been
described\cite{9} else where and will be omitted here.

The crystal structure of MgB$_{2}$ is illustrated in Fig. 1. It has a
hexagonal unit cell, with the primitive vectors $a=3.086$ \AA\ and $c=3.504$
\AA\ respectively.\cite{10} The space group for this compound is {\it P6/mmm}
(No. 191). From the figure is possible to notice that the boron atoms are
arranged in layers, with the next layers of Mg interleaved between them. In
order to form the unit cell two layers of magnesium and one of boron are
needed along the c-axis in the hexagonal configuration. For the basis set
for magnesium and boron, we selected 6-21G* and 6-21G basis set
respectively, provided in Ref. (11). The selection of the basis set was
obtained between a compromise between accuracy and computational resources.

\section{ RESULTS AND DISCUSSION}

For this calculation, we used the experimental lattice parameters provided
by Nagamatsu {\em et al.}\cite{1} of a= 3.086 \AA , c= 3.524 \AA\
respectively. Band structure calculations for MgB2 are depicted in Fig. 2.
In order to obtain the band structure for this compound, we used 133 {\em k}%
-points sampling the FBZ.

Our results are in good qualitative agreement with the results obtained by
other theoretical methods, such as the results obtained by Kortus {\em et al}%
.\cite{6} who used a general potential LAPW code, whilst Satta {\em et al}. 
\cite{12} using local-density approximation to the density functional
theory, and full-potential linearized augmented plane waves, obtained
similar results. Furthermore, Belaschenko\cite{13} calculated the electronic
structure for this compound using Stuttgard LMTO-TB (ASA) to compute the
energy bands with similar results.

We spanned the FBZ covering the three dimensional space going from $\Gamma $%
-M-K-$\Gamma $-A-L-H-A of $k_{x}=p/a,k_{y}=p/b,k_{z}=p/c$. The $\Gamma $-M-K-%
$\Gamma $ lines are in the basal plane, while the A-L-H-A are located on the
top of the plane at $k_{z}$. The upper part of the valence band for MgB$_{2}$%
, composed of B 2p-states which form two peculiar set of bands of $\sigma $%
(2p$_{x,y}$) and $\pi $(p$_{z}$) character, whose $k$ dependence changes
considerably while spanning the three dimensional space. For B 2p$_{x,y}$
the most pronounced dispersion is along $\Gamma $-K. These bands are of
quasi-two dimensional type and form an almost flat zone along $\Gamma $-A,
which reflects the distribution of pp$\sigma $-states in the B layers. These
states provide a big contribution to the density of states close to the
Fermi energy, providing the metallic properties to the borides.

On the other hand, B p$_{z}$ like bands are responsible to the weak pp$\pi $
interactions. These bands are considered to be a three dimensional like
type, which have maximum dispersion along $\Gamma $-A. Furthermore, Mg s, p
and B s states are admixed with B 2p bands close to the bottom of the
valence band, as well as in the conduction band. Hence the electronic
properties of MgB$_{2}$ are associated with metallic 2p-states of B atoms
located in planar nets, which determines the DOS close to the Fermi level.

For the total and projected density of states (PDOS) for B and Mg, see Fig.
3. Concentrating on those states close to the Fermi level, which presumably
are responsible for the superconductor behavior of the material, as we
mentioned before in the energy band discussion, the main contributions comes
from p-$\sigma $ bonding and p$_{z}$ having $\pi $ bonding and anti bonding
character on the basal and top planes, respectively. Which one of these two
bands is responsible for the superconductor behavior is unclear so far.

In summary, we have shown energy bands, as well as total and projected
density of states for the medium temperature superconductor MgB$_{2}$. From
these analysis, we were able to infer the most important contributions from
each atom to the superconductor behavior for the compound. So far, this
study is not conclusive, but continues in order to provide other important
physical properties such as charge density profiles and Mulliken population
analysis.

\acknowledgments

Both of us acknowledge DGSCA-UNAM and CoNaCyT for providing financial
support.

\newpage

\begin{figure}[tbp]
\caption{Crystal structure of MgB$_{2}$. Large spheres represent magnesium
and small spheres represent boron atoms.}
\label{fig1(a)}
\end{figure}

\begin{figure}[tbp]
\caption{Crystal structure of MgB$_{2}$. Large spheres represent magnesium
and small spheres represent boron atoms.}
\label{fig1(b)}
\end{figure}

\begin{figure}[tbp]
\caption{Band structure calculations for MgB$_{2}$.}
\label{fig.2}
\end{figure}

\begin{figure}[tbp]
\caption{Total density of states for MgB$_{2}$.}
\label{fig.3(a)}
\end{figure}

\begin{figure}[tbp]
\caption{Boron projected DOS for MgB$_{2}$.}
\label{fig.3(b)}
\end{figure}

\begin{figure}[tbp]
\caption{Magnesium projected DOS for MgB$_{2}$.}
\label{fig.3(c)}
\end{figure}

\end{document}